# A Novel Method for Spectrum Sensing of Linear Modulation Schemes

A. K. Karthik[1], Jameer Ali M.S[2], Mohammed Zafar Ali Khan[3], A. Bhagavathi Rao[4]
UURMI systems Pvt. Ltd., Door No: 860/A, Road No: 45,
Jubilee Hills, Hyderabad, A.P., India
[1]krishnakarthika@uurmi.com
[2]jameeralim@uurmi.com
[3]zafar@iith.ac.in
[4]bhagavatrao123@gmail.com

*Abstract*:
*In this paper, we propose and evaluate a novel algorithm for performing spectrum sensing on linear modulations based on second order cyclic features of the received signals. The proposed approach has similar computational complexity to that of energy detection, and outperforms energy detection and other sensing schemes such as Eigenvalue based sensing in the presence of noise uncertainties for a given value of probability of false alarm.*

*Keywords—Cyclostationary based detection, eigen-value base sensing, electronic warfare, spectrum sensing,*

## I. INTRODUCTION

Electronic warfare (EW) refers to any action involving the use of the electromagnetic spectrum or directed energy to control the spectrum, attack an enemy, or impede enemy assaults via the spectrum. The purpose of electronic warfare is to deny the opponent the advantage of, and ensure friendly unimpeded access to, the EM spectrum. Sensing, classifying, and adapting signals are an important aspect of Electronic warfare and Cognitive Radio. A fast and robust spectrum sensing approach is critical for EW.The basic idea behind spectrum sensing is tosense the signal in the presence of noise.Many approaches addressing the problem of spectrum sensing such as Energy detection, Match filter based sensing, Cyclostationary-based sensing, Eigen-value based sensing etchave been proposed in the literature [1]. We describe some of the prominent approaches used in spectrum sensing.

The Energy detection approach is a simple method where the signal is declared to be present if the total energy of the received signal is greater than a certain threshold [2]. This approach, however, is very sensitive to knowledge of the noise variance, and the performance degrades significantly when the noise variance is not known perfectly. Also it is also not optimal for detecting correlated signals, which is the case for most practical applications [3]. The Matched filter detector is an optimal detector since it maximizes the received signal-to-noise ratio (SNR) in communication systems [4], and in [5], it is shown that waveform-based sensing outperforms energy detector based sensing in reliability and convergence time. Furthermore, it is shown that the performance of the sensing algorithm increases as the length of the known signal pattern increases. However, a significant drawback of the matched filter based sensing is that it needs the prior knowledge of the primary user's signal such as the modulation type and order, pulse shaping and packet format and it is also susceptible to synchronization errors. The Cyclostationarity based approaches [6]--[9] for spectrum sensing exploits the cyclic properties that are present in communication signals, but absent in noise which exhibits W.S.S (wide sense stationarity). In Eigen value based sensing approaches [10], the ratio of the maximum to the minimum eigenvalue is calculated and if the value is greater than the threshold, the signal is present; else the signal is absent. This method is robust to noise uncertainties, and can even perform better than Energy detection, when the signals to be detected exhibit high degree of correlation.

In this paper, we propose and evaluate a new approach for spectrum sensing, by exploiting the Cyclostationarity in the received signal. Unlike the traditional approaches which use the statistical tests [11] to determine the presence of cyclostationarity, the proposed approach doesn't explicitly require any test to determine cyclostationarity. We perform sensing using the statistics of the Fourier series coefficient rather than use the statistical tests. The proposed approach is also computationally simpler than the typical Cyclostationarysensing schemes.

The rest of this paper is organized as follows: In Section II, we present the signal model along with the considered problem statement. In Section III, we briefly present some information regarding second-order cyclostationarity. Performance evaluation and comparisons are given in Section IV, and finally, our conclusions are offered in Section V.

## II. SIGNAL MODEL AND PROBLEM STATEMENT

### 1. Signal Model

We consider two hypotheses $H_0$ and $H_1$ which are described as follows

$$\begin{array}{lll} y(n) = & x(n) + w(n) & H_1 \\ y(n) = & w(n) & H_0 \end{array} \quad (1)$$

Where $x(n) = x(t)|_{t=nT_s} = \sum_{k=-\infty}^{\infty} s_k g(nT_s - kT)$ is a linear modulated signal, where $T_s$ represents the sample time, $T$ represents the symbol time,$s_k$ represents the digital symbols, $g(.)$ represents the pulse shape used in the transmitter. We assume the symbols are zero mean and uncorrelated i.e., $E[s_k] = 0$, $E[s_k s_m^*] = \delta_{km}$, where $\delta_{km}$ represents the Kronecker delta function.$w(n)$represents the white Gaussian noiseof mean 0 and variance $\sigma_w^2$. The probability of false alarm is denoted by $P_f = P(H_1/H_0)$, the probability of detection is denoted by $P_d = P(H_1/H_1)$,

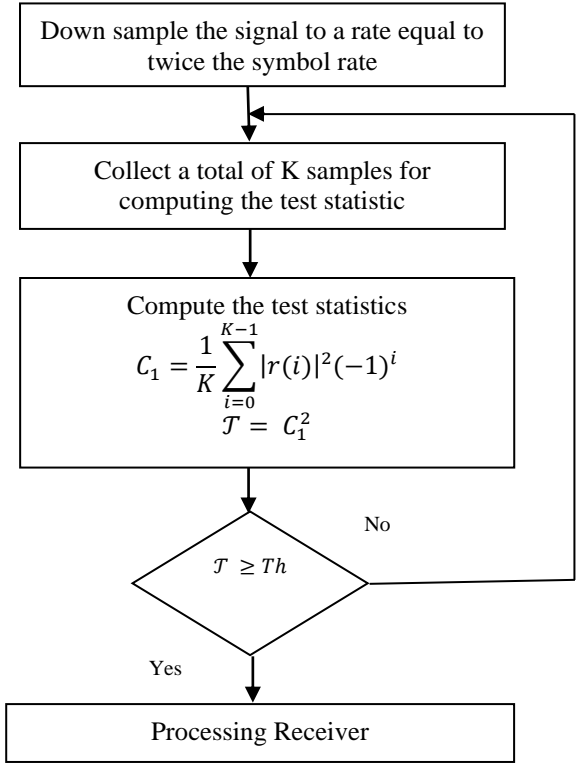

Figure 1. Flow Diagram of the Proposed Algorithm

and the probability of miss is denoted by $P_m = (1 - P_d)$.

*2. Problem Statement*

In this paper we address the problem of sensing the signal of linear modulated signal $x(n)$ from the received samples $y(n)$, for a fixed value of $P_f$. The main aim is to maximize the value of $P_d$ for a given value of $P_f$.

### III. CYCLIC AUTO-CORRELATION FUNCTION

The second order cyclic auto-correlation function $R_y^\alpha(\tau)$ of $y(t)$ at the cycle frequency α is defined as

$$R_y^\alpha(\tau) = \lim_{Z \to \infty} \frac{1}{Z} \int_{-Z/2}^{Z/2} y(t+\tau) y^*(t) e^{-j2\pi\alpha t} dt \quad (2)$$

For the considered linear modulated signals, the cycle frequency is observed at $\alpha = \frac{K}{T}, K \in \mathbb{Z}$. An estimate of the second order cyclic auto-correlation function can be calculated from the received samples $y(iT_s)$ using (3)

$$\hat{R}_y^\alpha(lT_s) = \frac{1}{K} \sum_{n=0}^{K-1} y(n+l) y^*(n) e^{-j2\pi\alpha n} \quad (3)$$

Where `$K'$` represents the number of samples used for detection. As we observe from (3), the cycle frequency at $\frac{K}{T}$ in the continous time domain is mapped to $\frac{KT_s}{T}$ in the discrete domain.

### IV. CONSTANT FALSE-ALARM RATE (CFAR) DETECTORS

The performance of detector is determined by the probability of detection $P_d$. We use the Neyman-Pearson test to evaluate and compare the performance of various spectrum sensing approaches. In this test, we fix the threshold λ to achieve a given value of $P_f$, and evaluate $P_d$ for various SNRs using the same threshold λ. Based on the comparison of test statistic, $\mathcal{T}$ with threshold λ, we have

$$P_f = P(\mathcal{T} > \lambda; H_0)$$
$$P_d = P(\mathcal{T} > \lambda; H_1)$$

The threshold λ is chosen such that for a given value of $P_f$, we can maximize $P_d$. In our paper, we compare the performance of our approach to various approaches available in the literature using the CFAR test.

### V. PROPOSED APPROACH

In our approach, we exploit the fact the linearly modulated signals exhibit cyclostationarity. We perform spectrum sensing on the linear modulation at 2 samples/symbols, i.e., the discrete cycle frequency is at $\pm \frac{1}{2}$. The test statistic $\mathcal{T}$ is defined as follows

$$\mathcal{T} = \left| \frac{1}{K} \sum_{n=0}^{K-1} |y(n)|^2 (-1)^n \right|^2 \quad (4)$$

Where $K$ represents the number of samples used for estimation.

*1. Probability of false alarm*

We now present the basic steps to determine the $P_f$ for the test statistic given above. We first define

$$C_1 = \frac{1}{K} \sum_{n=0}^{K-1} |y(n)|^2 (-1)^n \quad (5)$$

$$\mathcal{T} = C_1^2 \quad (6)$$

Under $H_0$, $P(\mathcal{T}/H_0)$ has $\chi^2$ distribution with 1 degree of freedom with mean $\mu_0 = \frac{\sigma_w^4}{K}$ and variance $\sigma_0^2 = 2\frac{\sigma_w^8}{K^2}$. We evaluate the value of $P_d$ using simulations and compare the performance of the proposed approach to other approaches available in the literature.

### VI. ALGORITHM FOR SPECTRUM SENSING

We now present the basic steps of our proposed sensing algorithm.

- **STAGE-1**: In this stage, we reduce the sample rate of the received signal to 2 samples/symbol.
- **STAGE-2**: After collecting $K$ samples, we compute the test statistic $\mathcal{T}$ as described in (4).
- **STAGE-3**: Compute the threshold value λ for a given value of $P_f$ from the central chi-square tables.
- **STAGE-4**: If $\mathcal{T} \geq \lambda$, then declare that the signal is present, else declare that noise is present.

### VII. SIMULATION RESULTS

*1. Simulation Setup*

We have peroformed simulations to evaluate the performance of the proposed algorithm. We have considered $x(n)$ as BPSK modulated signal, the pulse shape $g(.)$ is chosen as a Gaussian pulse shaping filter with $BT = 0.5$. We fix the value of $P_f = 0.1$. We present our results in terms of probability of detection, $P_d$ with respect to SNR. We compare the performance of our approach with Energy detection and Eigen value based sensing approaches.

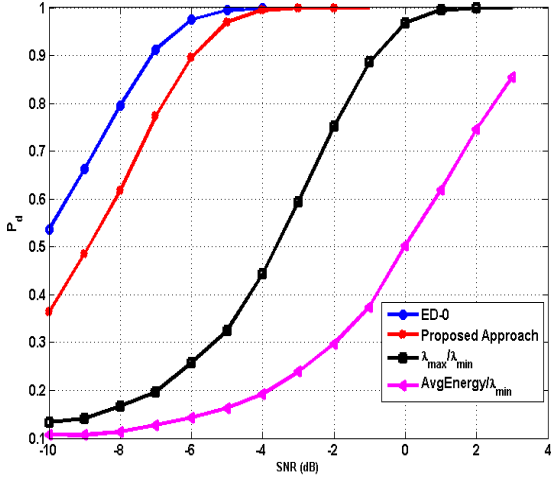

Figure 2. Probability of detection under perfect knowledge of noise

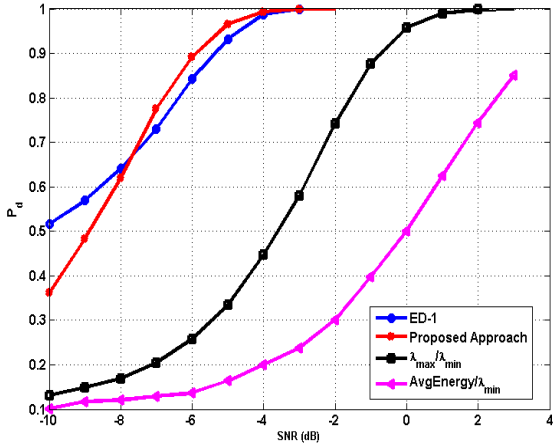

Figure 3. Probability of detection under ±1 dB noise uncertainity

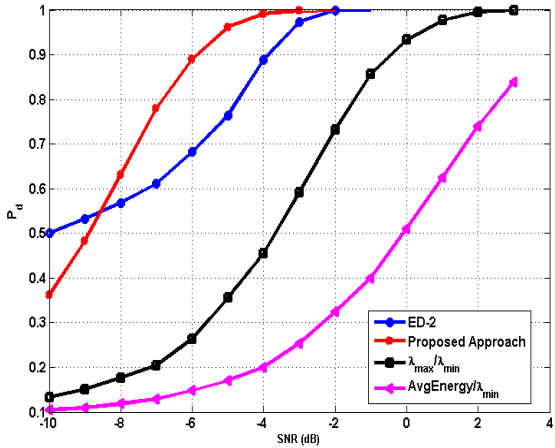

Figure 4. Probability of detection under ±2 dB noise uncertainty

In the case of energy detection, the test statistic is defined as

$$C_0 = \frac{1}{K}\sum_{n=0}^{K-1}|y(n)|^2$$

| Scheme | $P_f$ | $P_d$ (SNR = -12 dB) | $P_d$ (SNR = -10 dB) | $P_d$ (SNR = -8 dB) |
|---|---|---|---|---|
| ED – 0 dB | 0.1015 | 0.9288 | 0.9984 | 1.0000 |
| ED – 1 dB | 0.4450 | 0.5704 | 0.6508 | 0.7861 |
| ED – 2 dB | 0.4663 | 0.5425 | 0.5739 | 0.6429 |
| CD – 0 dB | 0.0999 | 0.7931 | 0.9834 | 1.0000 |
| CD – 1 dB | 0.1068 | 0.7904 | 0.9823 | 0.9999 |
| CD – 2 dB | 0.1202 | 0.7960 | 0.9785 | 0.9998 |
| MME – 0 dB | 0.0954 | 0.2440 | 0.4639 | 0.8441 |
| MME – 1 dB | 0.0996 | 0.2473 | 0.4827 | 0.8348 |
| MME – 2 dB | 0.0981 | 0.2567 | 0.4979 | 0.8039 |
| EME – 0 dB | 0.0932 | 0.1482 | 0.2510 | 0.4834 |
| EME – 1 dB | 0.0952 | 0.1490 | 0.2578 | 0.5016 |
| EME – 2 dB | 0.0957 | 0.1644 | 0.2760 | 0.5178 |

Table I. Comparison of probability of false alarm

Whereas in case of eigen value based sensing, the test statistics are defined as

$$\mathcal{T}_{MME} = \frac{\lambda_{max}}{\lambda_{min}}$$

$$\mathcal{T}_{EME} = \frac{c_0}{\lambda_{min}},$$

where $\lambda_{max}$ and $\lambda_{min}$ represent the maximum and minimum eigen value of the covariance matrix. We consider 1000 symbols for computing test statistics. The results for $P_d$ and $P_f$ are presented in the following sections.

*2. Observations*

We now present some observations from the achieved results.

- From Figure 2, we observe that the proposed approach performs worse than Energy detection, but performs better than the eigen value based approaches when the noise variance is known perfectly.
- From Figure 3, we observe that the proposed approach performs significantly better than Energy detection and the Eigen value based approachesunder noise uncertainities in the order of ±1dB.
- From Figure4, we observe that the proposed approach performs significantly better than Energy detection and the Eigen value based approaches under noise uncertainities even in the order of ±2dB.
- It is also observed that the performance of the proposed approach doesn't degrade in presence of noise uncertainties similar to the Eigen value based schemes.
- From the Table I, we see that even in the presence of noise uncertainties, the probability of false alarm does not increase significantly like Energy detection.(Similar to MME and EME).

As we can see from the results, the proposed approach outperforms both the Energy detection as well as the Eigen-value based sensing schemes in the presence of noise uncertainties even of the order of 1-2 dB. However, when the noise variance is known perfectly, it is worse than Energy detection by 1 dB for of $P_d = 0.9$.

VIII. CONCLUSION

The statistical test approaches for the general Cyclostationary tests are computationally complex while the proposed approach doesn't add significant computational complexity to the system. The statistical test approaches doesn't require any information regarding the noise variance, while the proposed approach requires knowledge of the noise variance. However, the performance of the approach is not affected to a great extent in the presence of noise uncertainties as shown in the simulations. The proposed method is computational simpler than EME and MME, and gives better results than EME and MME even in the presence of noise uncertainties.

## BIODATA OF AUTHORS

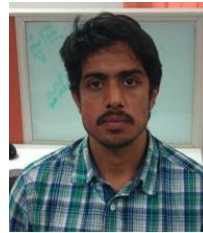
A.K.Karthik received his B.Tech in 2010, M.S (by research) in 2011 from IIIT-Hyderabad, India. He is presently working as a project engineer in UURMI systems Pvt. Ltd. His research interests are in communication system design, detection and estimation theory, blind modulation classification, blind deconvolution and cognitive radios.

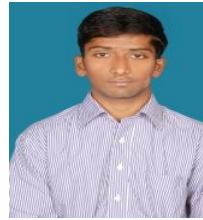
Jameerali M.S received his B.Tech from Jawaharlal Nehru Technological University, India in 2011. He is presently working as a project engineer in UURMI systems Pvt. Ltd. His research interests are in communication system design, detection and estimation theory, blind modulation classification and cognitive radios.

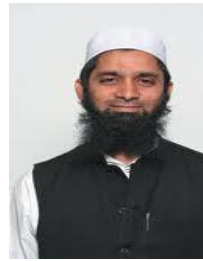
Dr.Zafar Ali khan received his B.E from Osmania University, Hyderabad, India, M.Techfrom IIT Delhi and Ph.D. from IISc,Bangalore. He is currently with IIT Hyderabad as an associate professor. His research interests are in coded modulation, Space-Time Coding, and Signal Processing for wireless Communications. He also author of book"Single and double symbol decodable Space time block codes".

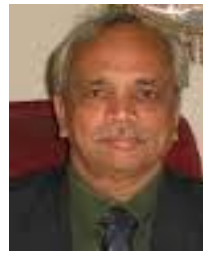
Mr. Bhagavathi Rao is a retired scientist from DLRL, Hyderabad. He has served as director of electronics, R&D headquarters, Delhi and also as the CEO of GATECH, Hyderabad, and group head of electronic warfare systems at DLRL among other assignments. He is presently associated with Uurmi Systems guiding the advanced development of next generation communication and radar systems.